\documentclass[secnumarabic,amssymb, nobibnotes, aip, reprint]{revtex4-1}

\usepackage{graphicx}
\usepackage{amsmath,times}

\def\U#1{{\rm #1}}
\def\vct#1{\mbox{\boldmath $#1$}}
\def\sub#1{_{\rm #1}}

\begin{document}

\title{Storage and retrieval of electromagnetic waves 
using electromagnetically induced transparency in a nonlinear metamaterial}

\author{Toshihiro Nakanishi}
\email[E-mail: ]{t-naka@kuee.kyoto-u.ac.jp}
\author{Masao Kitano}
\affiliation{{Department of Electronic Science and Engineering,
Kyoto University, Kyoto 615-8510, Japan\nolinebreak[4]}}
\date{\today}%

\begin{abstract}
 We investigate the storage and retrieval of electromagnetic waves using
 a nonlinear metamaterial, analogous to the electromagnetically induced
 transparency (EIT) observed in atomic systems.
 We experimentally demonstrate
 the storage of the electromagnetic wave by reducing an
 auxiliary ``control'' wave;
 the stored wave is then released by recovering the control wave.
 We also confirm that the metamaterial can store and reproduce the phase
 distribution of the original input wave.
 These effects confirm a remarkable analogy between the metamaterial and
 an atomic EIT medium. 
\end{abstract}

\maketitle

Electromagnetically induced transparency (EIT) is a nonlinear optical
phenomenon whereby an opaque medium is made transparent for a probe light
by the incidence of
an auxiliary ``control'' light in an
extremely narrow bandwidth \cite{Harris1997,Fleischhauer2005}.
In the frequency regions with high transparency,
the group velocity of the probe light is dramatically
reduced.
Characteristic features of EIT, including a narrow-band transparency and slow propagation,
have been extensively investigated in quantum systems composed of
three-level atoms.
Slow-light propagation in an EIT medium can be exploited for
``optical memory'' applications, or for storing and
retrieving light by dynamically modulating the control light,
whose intensity is proportional to
the group velocity of the probe light \cite{Fleischhauer2000,Phillips2001,Liu2001,Turukhin2001}.

Narrow-band transparency  and slow propagation can also be realized in other
systems such as an optical waveguide coupled with optical cavities
\cite{Chu1999,Xu2006,Totsuka2007} and
optomechanical systems\cite{Lin2010a,Weis2010,Safavi-Naeini2011},
 to name but a few \cite{Yoo2014, GarridoAlzar2002}.
Metamaterials composed of artificially designed structures
can also mimic the EIT effects observed in atomic systems.
Since the experimental demonstration of the EIT-like effect in the microwave
frequency regime \cite{Fedotov2007,Papasimakis2008},
various types of the EIT-like metamaterials have been reported to operate
in different frequency ranges \cite{Zhang2008a,
Liu2009,Hokari2014,Yang2014}.
The EIT-like metamaterials can serve various applications, including
high-accuracy sensing \cite{Lahiri2009,Liu2010a},
nonreciprocal propagation \cite{Mousavi2014,Floess2017}, and
nonlinearity enhancement\cite{Yang2015}.
For more practical applications,
property-tunable metamaterials have been realized
using nonlinear diodes \cite{Meng2014b},
superconductors \cite{Kurter2011, Limaj2014},
air-discharge plasma \cite{Tamayama2015a},
photocarrier excitation in semiconductors \cite{Gu2012,Miyamaru2014}, and 
microelectromechanical systems \cite{Pitchappa2016}.

The storage of electromagnetic waves in a
metamaterial requires the rapid control of the group velocity.
The first demonstration of electromagnetic-wave storage was
realized using an EIT-like metamaterial with variable capacitors \cite{Nakanishi2013}.
The metamaterial had similar functionality to the original EIT medium
with three-level atoms, but the transparency or the group velocity
was controlled by a bias voltage applied to the variable capacitors,
rather than by the electromagnetic waves.
In this sense, the effect can be regarded as displaying ``static-electric-field-induced transparency.''
The bias voltage on each constituent, or ``meta-atom,'' should be applied via a bias
circuit because a static field cannot propagate in free space.
This constitutes a critical obstacle to 
increasing the number of meta-atoms or extending the frequency range.

Here, we focus on a metamaterial showing genuine EIT effect,
whereby transparency results from the incidence of an auxiliary electromagnetic
wave (the control wave).
We previously reported
the static control of the EIT effect by using control waves
via three-wave mixing in varactor diodes included within the meta-atoms \cite{Nakanishi2015}.
In this paper, we demonstrate the storage and retrieval of
electromagnetic waves in the true EIT
metamaterial by dynamically modulating the control waves
in the same way as in the original atomic EIT systems.

First, we review a conventional method used to mimic 
the EIT-related phenomena, namely a narrow-band transparency and slow propagation.
EIT-like metamaterials are well described using a coupled-resonator
model \cite{GarridoAlzar2002}.
It is intriguing that Fano resonances \cite{Fano1961, Miroshnichenko2010} can also be explained by
this classical model \cite{Joe2006, Satpathy2012},
because the EIT effect can be regarded as a special case of Fano
resonances, where the spectral lineshape becomes symmetric.
One of the resonant modes, called a radiative mode, is responsible for 
directly interacting with propagating electromagnetic waves.
The other resonant mode, called a trapped mode, is uncoupled from
the external field and is excited only via coupling with the
radiative mode.
If the resonant frequencies of the radiative and trapped modes
are close to the frequency of the incident waves
and there is some degree of coupling between them,
the waves received through the radiative mode are effectively
transferred to the trapped mode.
The electromagnetic waves are temporarily stored in the trapped mode
and are subsequently re-emitted through the radiative mode.
Owing to the high quality factor of the trapped mode,
dissipation and scattering are substantially suppressed.
In addition, the temporary storage in the trapped mode causes a 
propagation delay. Consequently, transparency and slow propagation in
the metamaterial are realized.
If the coupling between two resonant modes
depends on a controllable parameter,
the energy flow between the radiative and trapped mode
can be dynamically controlled.
By removing the coupling during the interaction
with the external waves, the energy in the trapped mode is
completely captured in the metamaterial,
and thus, storage of electromagnetic waves can be achieved\cite{Nakanishi2013}.

\begin{figure}[t]
 \begin{center}
  \includegraphics[scale=0.43]{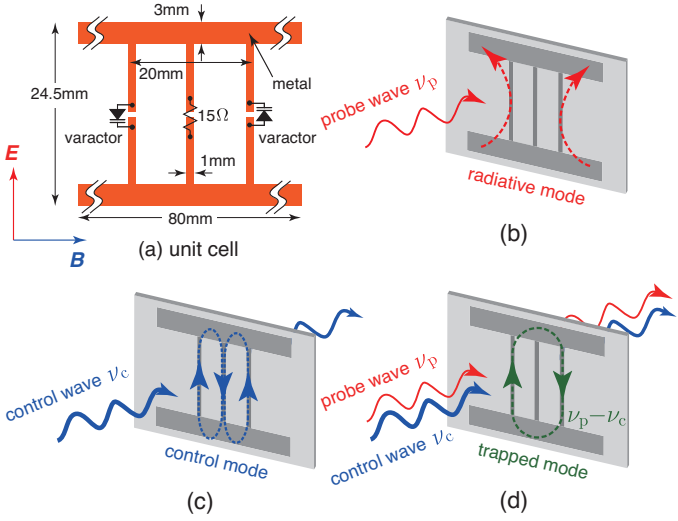}
  \caption{(a) Structure of a meta-atom and current flow for the (b) radiative, (c) control, and (d) trapped modes.}
  \label{structure}
 \end{center}
\end{figure}
Next, we introduce a metamaterial showing genuine EIT, as
realized in atomic EIT systems.
The metamaterial design, shown in Fig.~\ref{structure}(a),
is the same as in our previous work \cite{Nakanishi2015}.
The directions of the electric field $\vct{E}$ and magnetic field $\vct{B}$
of the incident waves are also displayed.
The dimension of the unit structure is 120\,mm $\times$ 25\,mm.
The structure, made of copper, is fabricated on a dielectric
material of thickness 0.8\,mm and permittivity of 3.3.
Two nonlinear capacitors, called varactor diodes,
whose capacitances are functions of the voltage across the diode,
are inserted with opposite directions on the two lateral circuit paths.
The metamaterial has three resonant modes: a radiative mode,
a trapped mode, and an additional mode called the control mode.
The current flow for these modes is illustrated using the dashed lines with
arrows in Figs.~\ref{structure} (b)--(d).
The functions of the radiative and trapped modes are the same
as those in the conventional EIT-like metamaterials discussed above,
but the resonant frequencies, $f\sub{r}$ for the radiative mode
and $f\sub{t}$ for the trapped mode, are notably different.
The probe wave (with frequency $\nu\sub{p}$)
and the control wave (with frequency $\nu\sub{c}$)
can directly excite the radiative mode (Fig.~\ref{structure}(b))
and the control mode (Fig.~\ref{structure}(c)),
respectively.
On the other hand, the trapped mode cannot be excited by an incident wave with one
frequency only.
The transmission of the control wave can be made relatively strong
by adjusting $\nu\sub{c}$ to be located in the tail of the 
control-mode resonance at $f\sub{c}$.

In the absence of the control wave,
the transmission of the probe wave (with $\nu\sub{p}=f\sub{r}$)
is low because of the high radiation loss of the radiative mode.
On the other hand, in the presence of an intense control wave,
the probe wave becomes mixed with the control wave 
to produce other components with the frequencies $|\nu\sub{p}\pm \nu\sub{c}|$ via
three-wave mixing in the nonlinear capacitors.
If the resonant frequency of the trapped mode $f\sub{t}$ satisfies
the condition $f\sub{t}=\nu\sub{p} + \nu\sub{c}$ or
$f\sub{t}=\nu\sub{p} - \nu\sub{c}$,
the energy received in the radiative mode becomes effectively transferred
into the trapped mode owing to the resonance of the generated
components. (The proposed metamaterial is designed to satisfy $f\sub{t}=\nu\sub{p} - \nu\sub{c}$.)
The reverse process also occurs, and the radiative and trapped modes
become coupled via three-wave mixing with the control wave.
As a result, the metamaterial becomes transparent for the probe wave.

\begin{figure}[b]
 \begin{center}
  \includegraphics[scale=0.7]{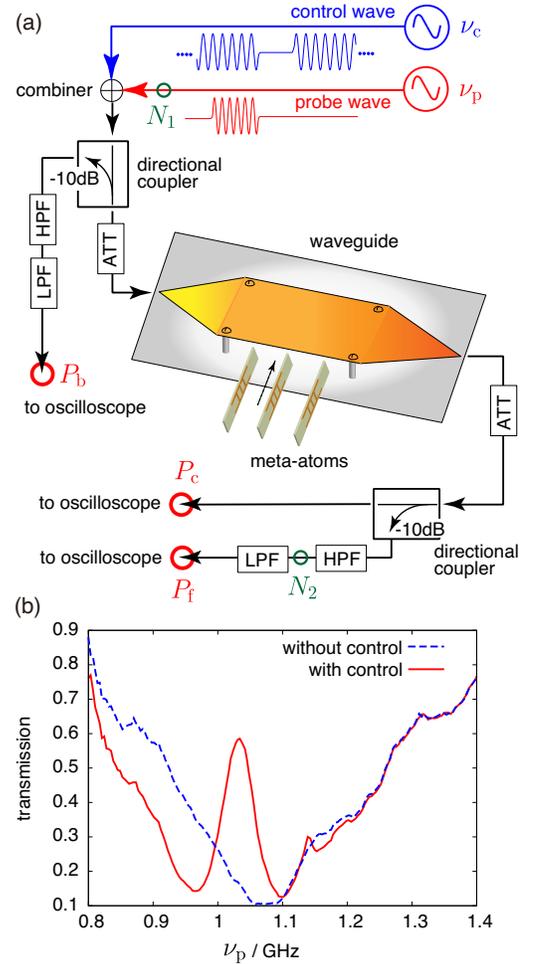}
  \caption{(a) Overview of the experimental setup. Directional couplers
  pick up 10\% of the transmitted or reflected signal.
  ATT: 5dB attenuator.
  LPF: low-pass
  filter with an insertion loss of $10\,\U{dB}$ at $1.07\,\U{GHz}$.
  HPF: high-pass filter with an insertion loss of $10\,\U{dB}$ at
  $0.865\,\U{GHz}$. (b) Transmission spectra for the probe wave in the presence or absence of the control wave.}
  \label{setup}
 \end{center}
\end{figure}

Figure \ref{setup}(a) shows a schematic diagram of the experimental
setup for performing the transmission measurements and for demonstrating the storage of
electromagnetic waves.
The resonant frequencies of the radiative, trapped, and
control modes are $f\sub{r} = 1.05\,\U{GHz}$, $f\sub{t}=0.65\,\U{GHz}$,
$f\sub{c}=0.4\,\U{GHz}$,
respectively \cite{Nakanishi2015}.
The probe and control waves are combined and sent to an
open-type waveguide, in which the meta-atoms are placed.
The waveguide, designed as a stripline with a height of $25\,\U{mm}$ and a
width of $122\,\U{mm}$, supports the propagation of a quasi-TEM mode
\cite{Nakanishi2012b}.
A unit comprising a directional coupler and a 5 dB attenuator is inserted at each
port of the waveguide.
The attenuators serve to suppress undesired multiple reflections.
We utilize the directional couplers followed by frequency filters,
which transmit the probe wave, in order to monitor the transmitted and
reflected signals for the probe wave.

First, we observed transmission spectra for the probe wave in the presence or absence of the control wave for a single-layer metamaterial.
A network analyzer was introduced to acquire spectra for the transmission between points $N_1$ and
$N_2$ in Fig.~\ref{setup}(a),
as the probe waves were swept from $0.8$ to $1.4\,\U{GHz}$.
The transmission spectra were normalized by a signal transmitted through the empty waveguide.
The transmission spectrum without the control wave is shown by
a dashed line in Fig.~\ref{setup}(b).
A broad transmission dip is observed around the resonant frequency of the radiative mode.
The solid line in Fig.~\ref{setup}(b) represents the transmission spectrum
in the presence of a continuous control wave with a fixed frequency of
$\nu\sub{c}=0.45\,\U{GHz}$,
which is located in the tail of the resonance of the control
mode centered at $f\sub{c}=0.4\,\U{GHz}$ and gives a transmission of 88\% (not
shown in
the figure).
The amplitude of the control wave is estimated to be $1.2\,\U{V}$ at the
waveguide input.
The observed spectrum clearly exhibits a sharp transparency window around $1.035\,\U{GHz}$.
In this case, the intense control wave connects the radiative
and trapped modes, and consequently induces EIT effect in the metamaterial.

\begin{figure}[b]
 \begin{center}
  \includegraphics[scale=0.5]{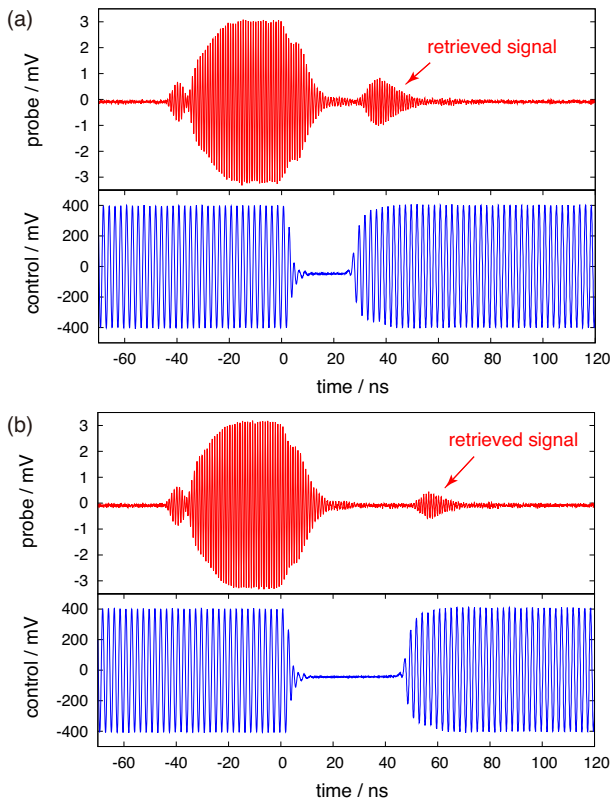}
  \caption{Storage and retrieval of electromagnetic waves, for a storage
  time of (a) $30\,\U{ns}$ or (b) $50\,\U{ns}$. The upper (lower) graph
  shows the probe (control) signal at port $P\sub{f}$ ($P\sub{c}$).}
  \label{exp}
 \end{center}
\end{figure}

Next, we describe procedures for storage and retrieval of electromagnetic waves,
which can be realized by dynamical modulation of the control wave.
The EIT effect is initially activated by illuminating the control
wave; the probe wave then travels slowly in the metamaterial.
During propagation, the control wave is switched off,
so that the probe wave is captured in the metamaterial.
This happens because the energy in the trapped mode can not escape 
via the radiative mode in the absence of three-wave mixing.
After some time period $\tau$, called the storage time,
the EIT effect is recovered by reintroducing the control wave.
Then, the energy stored in the trapped mode is released into the
waveguide through nonlinearity-assisted coupling with the radiative mode.
This procedures are notably identical to that for an
atomic EIT system.

By way of experimental demonstration,
we placed a three-layered metamaterial with a separation of $7.5\,\U{cm}$ in the waveguide.
The amplitude of the control wave, tuned to $\nu\sub{c}=0.45\,\U{GHz}$, could be
varied using an external controller within several nanoseconds.
We prepared a probe pulse of $30\,\U{ns}$ duration and carrier frequency $1.035\,\U{GHz}$,
which corresponds to the transparency window in the presence of the
control wave.
The amplitudes of the probe and control waves at the input port of the waveguide were
$260\,\U{mV}$  and $1.2\,\U{V}$, respectively.
Output signals were acquired using a high-speed oscilloscope connected
to ports $P\sub{c}$ and $P\sub{f}$ in Fig.~\ref{setup}(a).
The signal at port $P\sub{c}$ was used to monitor the control wave
in the absence of the probe wave,
and the signal at port $P\sub{f}$ (having passed through two filters to remove
the control wave and unnecessary harmonic waves) was used
to monitor the probe wave.
The timing of the control-wave modulation was adjusted
to allow the capture of the rear part of the probe pulse.

Figures \ref{exp}(a) and (b) show the observed signals for storage
times $\tau=30$ and $50\,\U{ns}$, respectively, which correspond to the
periods without the control wave.
The upper and lower graphs in each figure represent the time-domain waveforms
at ports $P\sub{f}$ and $P\sub{c}$, respectively.
The modulation of the control wave can be confirmed in each lower graph.
In both cases, the probe-wave signals were reduced when the
control waves were switched off.
This fact means that the probe waves were captured in the trapped mode,
which was decoupled from the radiative mode in the absence of the control waves.
After reintroducing the control waves, some
signals return, corresponding to the probe waves being released from the
metamaterial.
The decay rate of the retrieved signals
is determined by the lifetime of the trapped mode.
In the experiment, ohmic loss in the nonlinear capacitors
dominates.

If the metamaterial stores and retrieves the input pulse while preserving
the spatial phase distribution, the retrieved signal is released in the
forward direction; the backward wave can be suppressed.
In order to confirm this effect, or ``coherent'' storage,
we compared the retrieved waves propagating in the forward and backward directions
for single- and three-layered metamaterials.
The backward signal can be observed at port $P\sub{b}$ in Fig.~\ref{setup}(a).
The coupler and the filters introduced before port $P\sub{b}$
were the same as those for the forward wave,
and no calibration was necessary between the forward-propagating wave
at port $P\sub{f}$ and the backward-propagating wave at port $P\sub{b}$.

\begin{figure}[]
 \begin{center}
  \includegraphics[scale=0.45]{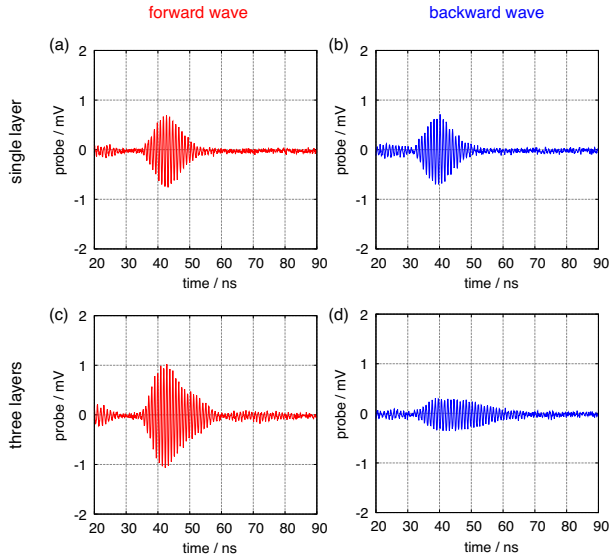}
  \caption{Signals from a single-layer metamaterial retrieved in the (a)
   forward or (b) backward directions,
  and those from three-layered metamaterial retrieved in the (c)
  forward or (d) backward directions.
  The storage time is $40\,\U{ns}$.}
  \label{fb}
 \end{center}
\end{figure}

The experimentally observed retrieved signals for the storage time $\tau=40\,\U{ns}$
are shown in Fig.~\ref{fb}.
The upper (lower) graphs show the results for the single-layer (three-layered)
metamaterial.
The graphs in the left (right) column correspond to the forward (backward) waves
observed at $P\sub{f}$ ($P\sub{b}$).
For the single-layer metamaterial, the amplitudes of the retrieved signals
observed in the forward and backward directions are
almost the same, as shown in Figs.~\ref{fb}(a) and (b),
because a single meta-atom releases the stored
energy in both directions with equal intensity.
For the three-layered metamaterial,
the retrieved signal in the forward direction (Fig.~\ref{fb}(c)) is enhanced,
while that in the backward direction (Fig.~\ref{fb}(d)) is suppressed.
This result is attributed to the interference in the radiation emitted from the
multiple meta-atoms.
The fact proves that the metamaterial stores and reproduces the probe wave
while preserving its phase distribution.
The interference in the experiment is not complete,
because of the insufficient number of meta-atoms,
incomplete transparency due to ohmic loss,
and inhomogeneity in the control wave.

In conclusion, we have experimentally demonstrated the storage and retrieval
of electromagnetic waves by exploiting EIT effect, implemented with a
nonlinear metamaterial designed for operation in the microwave
region.
The storage of the probe wave was achieved by switching the control wave off during the
propagation,
and the retrieval by recovering the control wave.
We also confirmed that the coherence was preserved in the storage and
retrieval processes
by comparing the retrieved signals released in the forward 
and backward directions.
These facts indicate that the metamaterial has the same ability
as the atomic EIT medium.
In principle, it should be possible to store the whole pulse
more efficiently by increasing the number of meta-atoms
and by reducing the undesired dissipation mainly caused by ohmic loss in the
nonlinear capacitors.
The nonlinear diodes can operate only at low frequencies (e.g., 
microwaves)
but the extension to higher frequencies (e.g., the optical region)
can be made possible by incorporating other nonlinear optical effects into
the metamaterial.

The present research was supported by Grants-in-Aid
for Scientific Research No.~17K05075 and No.~16K13699.

\bibliographystyle{apsrev4-1}


%

\end{document}